\documentclass{jfm}
\input{Preamble}
\title[Sensitivity analysis of a thermo-acoustic system via adjoint equations]{Sensitivity analysis of a time-delayed thermo-acoustic system via an adjoint-based approach\thanks{This is a pre-print version. Published in J.\ Fluid\ Mech., vol. 719, 2013. Cambridge University Press$^{\copyright}$. DOI: http://dx.doi.org/10.1017/jfm.2012.639}}
\author[L. Magri, M.P. Juniper]
{Luca Magri \thanks{Email address for correspondence: lm547@cam.ac.uk} and Matthew P. Juniper}

\affiliation{Department of Engineering, University of Cambridge,\\
Trumpington Street, Cambridge, CB2 1PZ, U.K.}

\pubyear{xxxx}
\volume{xxx}
\pagerange{xxx--xxx}
\date{?? and in revised form ??}
\begin{document}
\maketitle
\begin{abstract}
We apply adjoint-based sensitivity analysis 
to a time-delayed thermo-acoustic system:
a Rijke tube containing a hot wire.
We calculate 
how the growth rate and frequency of small oscillations about a base state
are affected
either by a generic passive control element in the system (the structural sensitivity analysis)
or by a generic change to its base state (the base-state sensitivity analysis).
We illustrate the structural sensitivity 
by calculating the effect of a second hot wire
with a small heat release parameter. 
In a single calculation,
this shows how the second hot wire 
changes the growth rate and frequency of the small oscillations,
as a function of its position in the tube.
We then examine the components of the structural sensitivity
in order to determine the passive control mechanism
that has the strongest influence on the growth rate. 
We find that 
a force
applied to the acoustic momentum equation
in the opposite direction to the instantaneous velocity
is the most stabilizing feedback mechanism.
We also find that its effect is maximized 
when it is placed at the downstream end of the tube.
This feedback mechanism could be supplied, for example, by an adiabatic mesh.
We illustrate the base-state sensitivity
by calculating the effects of small variations in 
the damping factor, 
the heat-release time-delay coefficient, 
the heat-release parameter,
and the hot wire location.
%
%Sensitivity analysis requires 
%the direct governing equations (linearized about a base state)
%and the adjoints of these equations to be calculated.
%%
%We calculate these adjoints via two different approaches:
%(1) discretising the analytically-derived adjoint equations ($\CA$),
%(2) deriving the adjoints of the discretized direct equations ($\DA$). 
%%
%For this system, in which the linearized direct equations are easily discretized,
%we find that the $\DA$ method is more accurate and more easily implemented than the $\CA$ method. 
%%
%Indeed, 
%the results of the $\DA$ method exactly match 
%those found by finite difference,
%which is an exact but expensive method. 
%
The successful application of 
sensitivity analysis to thermo-acoustics 
opens up new possibilities for the passive control of thermo-acoustic oscillations
by providing gradient information 
that can be combined with constrained optimization algorithms
in order to reduce linear growth rates.
\\

%\textbf{Key words:} acoustics, instability, adjoint equations, sensitivity, linear control.  
\end{abstract}

\section{Introduction}
\label{sec_int}
\par
\par
In a thermo-acoustic system, heat release oscillations couple with acoustic pressure oscillations. If the heat release is sufficiently in phase with the pressure, these oscillations grow, sometimes with catastrophic consequences. Using adjoint sensitivity analysis, we identify the most influential components of a thermo-acoustic system and quantify their influence on the frequency and growth rate of oscillations. This technique shows how a thermo-acoustic system should be changed in order to extend its linearly stable region. 
\par
Adjoint sensitivity analysis of incompressible flows was proposed by \citet{HILL-92} and developed further by \citet{GIANNETTI-07} in order to reveal the region of the flow that causes a von K\'{a}rm\'{a}n vortex street behind a cylinder. 
They used adjoint methods to calculate the effect that a small control cylinder has on the growth rate of oscillations, as a function of the control cylinder's position downstream of the main cylinder. 
This control cylinder induces a force in the opposite direction to the velocity field. \citet{GIANNETTI-07} and \citet{GIANNETTI-10} considered this feedback only on the perturbed fields but \citet{MARQUET-08} extended this analysis to consider the cylinder's effect on the base flow as well.
\citet{SIPP-10} provide a comprehensive review of sensitivity analysis for incompressible fluids and
\citet{CHANDLER-12} extend this analysis to low Mach number flows in order to model variable density fluids and flames. 
\par
The aim of this paper is to extend adjoint sensitivity analysis to a thermo-acoustic system, which has not been attempted before. 
We investigate the thermo-acoustic system described by \citet{BALASUB-08a} and \citet{JUNIPER-11}. 
This is an open-ended tube, 
through which air passes, 
and which contains a hot wire at a given axial location. 
One-dimensional acoustic standing waves in the tube modulate the air velocity at the wire,
which in turn modulates the heat transfer from the wire to the air,
which is modelled with a modified form of King's law \citep{HECKL-90, MATVEEV-03}. 
This heat transfer occurs at the wire's location
but is not instantaneous.
The time taken for the heat to diffuse 
to the bulk fluid 
is modelled as a time delay 
between the velocity fluctuations and the heat release fluctuations.
\par
The analysis consists of three main steps. 
Firstly, we study the system as an eigenvalue problem in the complex frequency domain.
Secondly, we derive two sets of adjoint equations from the linearized governing equations. 
Thirdly, we use the adjoint equations to perform both a structural sensitivity analysis and a base-state sensitivity analysis. 
The structural sensitivity analysis
quantifies the effect that feedback mechanisms have
on the frequency and growth rate of oscillations. 
This analysis relies on studying the effect of a perturbation to the governing equations,
which is known as a structural perturbation. 
There are several components of the feedback
and, in this paper, we calculate all of them. 
We then illustrate the structural sensitivity by 
considering the effect of feedback from a second hot wire. 
The base state sensitivity analysis
quantifies the effect of a change in the constant coefficients of the governing equations. 
It does not involve a feedback mechanism. 
The base state in this thermo-acoustic model is represented by four parameters: the damping factor, $\mathit{\zeta}$; the heat-release time-delay coefficient, $\mathit{\tau}$; the heat-release parameter, $\mathit{\beta}$, and the hot wire location, $\mathit{x_h}$. 
This shows us how to change these parameters in order to most stabilize the system. 
In addition, we can also calculate the location of the first hot wire that makes the system most sensitive to base-state modifications. 
In the final section we apply this analysis to the passive control of an unstable nonlinear system.

\section{Thermo-acoustic model}
\label{sec_mod_def}
The thermo-acoustic system examined in this paper is a horizontal Rijke tube containing a hot wire. It is governed by the following nonlinear time-delayed equations:
\begin{eqnarray}
\label{equ_gov_mom_ndim}
&&\frac{\partial u}{\partial t} + \frac{\partial p}{\partial x} = 0 ,\\
\label{equ_gov_enr_ndim}
&&\frac{\partial p}{\partial t} + \frac{\partial u}{\partial x}
+ \zeta p
- \dot{q}= 0,\\
&&\dot{q}=\frac{2}{\sqrt{3}}\beta \label{qqq1_NL}
\left(
\left|
\frac{1}{3} + u(t - \tau)
\right|^{\frac{1}{2}}
-
\left(
\frac{1}{3}
\right)^{\frac{1}{2}}
\right)
\delta(x - x_h).
\end{eqnarray}
where $\mathit{u}$, $\mathit{p}$ and $\mathit{\dot{q}}$ are the non-dimensional velocity, pressure, and heat-release rate, respectively.  
The hot wire is placed at $\mathit{x}=\mathit{x_h}$, which is modelled by the Dirac delta (generalized) function $\mathit{\delta(x-x_h)}$. 
The system has four control parameters: $\mathit{\zeta}$, which is the damping; $\mathit{\beta}$, which encapsulates all relevant information about the hot wire, base velocity, and ambient conditions; $\mathit{\tau}$, which is the time delay, and $\mathit{x_h}$, which is the position of the hot wire. The values of $\mathit{\beta}$, $\mathit{\tau}$, and $\mathit{x_h}$ are given in the figure captions along with the damping constants $\mathit{c_1}$ and $\mathit{c_2}$. 
In \S \ref{numdisc} we will explain how $\zeta$ is related to $c_1$ and $c_2$.
Eqs.~(\ref{equ_gov_mom_ndim})-(\ref{equ_gov_enr_ndim}) are derived from the Navier-Stokes and energy equations by assuming first-order acoustics, as explained in \citet{CULICK-71}.
The heat-release rate in eq.~(\ref{qqq1_NL}) is modelled with a modified form of King's law \citep{HECKL-90, MATVEEV-03}. 
Note that throughout this paper we define the heat-release parameter $\mathit{\beta}$ to be $\sqrt{3}/2$ times the heat-release parameter $\mathit{\beta}$ defined in \citet{JUNIPER-11}.
The heat-release term (\ref{qqq1_NL}) is linearized around a fixed point of the system, where $|u_h|\ll1$. 
In addition, eq.~(\ref{qqq1_NL}) is linearized also in time assuming that the time-delay coefficient is sufficiently small compared with the period of the highest Galerkin mode (\S \ref{numdisc}):
\begin{equation}
\dot{q}=\beta \label{qqq1}
\left(
 u
-
\tau \frac{\partial u}{\partial t}
\right)
\delta(x - x_h) .
\end{equation}
By substituting eq. (\ref{equ_gov_mom_ndim}) into eq. (\ref{qqq1}), we obtain an equivalent expression for the linearized heat-release law:
\begin{equation}
\dot{q}=\beta \label{qqq2}
\left(
 u
+
\tau \frac{\partial p}{\partial x}
\right)
\delta(x - x_h) .
\end{equation}
%. 
%
It is important to anticipate that, although eq. (\ref{qqq1}) is physically equivalent to (\ref{qqq2}), the systems of the linearized governing equations (\ref{equ_gov_mom_ndim})-(\ref{equ_gov_enr_ndim})-(\ref{qqq1}) and (\ref{equ_gov_mom_ndim})-(\ref{equ_gov_enr_ndim})-(\ref{qqq2}) will produce two different sets of adjoint equations (\S \ref{adjoints}).

%In this paper we will introduce a second hot wire, labelled with the subscript $c$, to analyse the structural sensitivity. 
% I DON'T THINK IT IS NECESSARY TO MENTION THIS HERE - THE READER WILL GET TO THIS SOON ENOUGH AND WE CAN EXPLAIN IT THEN. 
%
%
\section{Numerical discretization} \label{numdisc}
The partial differential equations (\ref{equ_gov_mom_ndim})-(\ref{equ_gov_enr_ndim})-(\ref{qqq1}), which govern the thermo-acoustic system,
are discretized into a set of ordinary differential equations
by choosing an orthogonal basis that matches the boundary conditions. 
This procedure is also known as the Galerkin method. 
The variables are expressed as:
\begin{equation}
\label{equ_gal_u}
u(x,t)  = 
\sum_{j = 1}^{N}
\eta_j(t) \cos(j \pi x), \;\;\;\;
p(x,t)  = 
- \sum_{j = 1}^{N}
\left(\frac{\dot{\eta}_j(t)}{j \pi}
\right)
\sin(j \pi x).
\end{equation}
The state of the system is given by the amplitudes of the Galerkin modes that represent velocity, $\mathit{\eta_j}$, and those that represent pressure, $\mathit{\dot{\eta}_j/j\upi}$. The state vector of the discretized system is the column vector $\boldsymbol{\chi} \equiv (\boldsymbol{u}, \boldsymbol{p})^T$, where $\boldsymbol{u} \equiv (\mathit{\eta_1, \ldots, \eta_N})^T$ and $\boldsymbol{p} \equiv (\mathit{\dot{\eta}_1/\upi, \ldots, \dot{\eta}_N/N\upi})^T$.
The discretized problem can be represented in matrix notation:
\begin{equation}
\frac{\mathrm{d} \boldsymbol{\chi}}{\mathrm{d} t} = \mathsfbi{\Gamma} \boldsymbol{\chi} . \label{D}
\end{equation}
where $\mathsfbi{\Gamma}$ is the $2\mathit{N}\times2\mathit{N}$ direct matrix  and $\boldsymbol{\chi}$ is the  $2\mathit{N}\times1$ state vector. 
The basis functions, $\cos(j \pi x)$ and $\sin(j \pi x)$, are the eigenfunctions of the undamped acoustic system without the heater. 
The direct matrix $\mathsfbi{\Gamma}$ is shown in appendix \ref{append} in eq. (\ref{gammaAPP}). Note that, when the system has $\mathit{N}$ Galerkin modes, it has $2\mathit{N}$ degrees of freedom.

The linearized equations in \S \ref{sec_mod_def} are valid for small $|\mathit{u}_h|$ and $\mathit{\tau}\ll\mathit{T_j}$, where $T_j=2/\mathit{j}$ is the period of the $\mathit{j^{th}}$ Galerkin mode, as explained in \citet{JUNIPER-11}.
The results are presented here for a system with 10 Galerkin modes (as for system C in \citet{JUNIPER-11}).
We checked modal convergence considering more Galerkin modes and found that 10 modes provide an accurate representation of the system, as discussed in \S \ref{fefb}.

At the ends of the tube, $\mathit{p}$ and $\mathit{\partial u}/ \mathit{\partial x}$ are both set to zero, which means that the system cannot dissipate acoustic energy by doing work on the surroundings. Dissipation and end losses are modelled by the damping parameter for each mode $\mathit{\zeta_j} = c_1 j^2 + \mathit{c_2 \sqrt{j}}$, where $\mathit{c_1}$ and $\mathit{c_2}$ are constants. 
Oscillations of higher Galerkin modes decay very rapidly if no mechanism drives them. 
This damping model was used in \citet{BALASUB-08b} and was based on correlations developed by \citet{MATVEEV-03} from models in \citet{LANDAU}.
\section{Adjoint operator} \label{adjoints}
In this section the \emph{adjoint operator} is defined. 
This definition is an extension over the time domain of the definition given by \citet{DENNERY}.
Let $\mathrm{L}$ be a partial differential operator of order $\mathit{M}$ acting on the function $\mathit{q(x_1,x_2,\ldots,x_K,t)}$, where $\mathit{K}$ is the space dimension, such that $\mathrm{L}\mathit{q(x_1,x_2,\ldots,x_K,t)}=0$. We refer to the operator $\mathrm{L}$ as the \emph{direct operator} and the function $\mathit{q}$ as the \emph{direct variable}.
The adjoint operator $\mathrm{L}^+$ and adjoint variable $\mathit{q^+}$ are defined via the \emph{generalized Green\rq{}s identity}:
\begin{equation}
\int_0^T \! \int_{V}\! \bar{q}^+\mathrm{L}q-q\left(\overline{\mathrm{L}^+q^+}\right)\mathrm{d}V \mathrm{d}t=\int_0^T \int_{S}\!\; \sum_{i=1}^{K} \left[ \frac{\partial}{\partial x_i} Q_i \left(q,\bar{q}^+\right) \right] n_i \mathrm{d}S \mathrm{d}t + \int_{V}Q_i\left(q,\bar{q}^+\right)\rvert_0^T\mathrm{d}V.\label{Greens_gen}
\end{equation}
where $\mathit{i}=1,2,\ldots,K$ and $\mathit{Q_i(q,\bar{q}^+)}$ are functions which depend bilinearly on $\mathit{q}$, $\mathit{\bar{q}^+}$ and their first $\mathit{M-1}$ derivatives. The complex-conjugate operation is labelled by an overline. The domain $\mathit{V}$, is enclosed by the surface $\mathit{S}$, for which $\mathit{n_i}$ are the projections on the coordinate axis of the unit vector in the direction of the outward normal to the surface $\mathrm{d}\mathit{S}$. The time interval is $\mathit{T}$. The adjoint boundary conditions and initial conditions on the function $\mathit{q^+}$ are defined as those that make the RHS in eq.(\ref{Greens_gen}) vanish identically on $\mathit{S}$, $\mathit{t}=0$ and $\mathit{t=T}$.

The adjoint equations can either be derived from the continuous direct equations and then discretized ($\CA$, discretization of the Continuous Adjoint) or be derived directly from the discretized direct equations ($\DA$, Discrete Adjoint).
%, as explained by \cite{CHANDLER-12}.
%
For the $\CA$ method (\S\ref{tw} and \S\ref{fff}), 
the adjoint equations are derived by
integrating the continuous direct equations by parts
and then applying Green\rq{}s identity (\ref{Greens_gen}).
They are then discretized with the Galerkin method (\ref{equ_gal_u}).
The appendices of \citet{JUNIPER-11} show the intermediate steps.
Two different sets of adjoint equations are derived here, shown in table \ref{tab_adj}.
The first set, $\CAone$, is obtained from
(\ref{equ_gov_mom_ndim})-(\ref{equ_gov_enr_ndim})-(\ref{qqq1})
and produces the discretized adjoint matrix (\ref{phi1}). 
The second set, $\CAtwo$, is obtained from 
(\ref{equ_gov_mom_ndim})-(\ref{equ_gov_enr_ndim})-(\ref{qqq2}) and produces the discretized adjoint matrix (\ref{phi2}).
The difference arises merely because the governing equations are arranged differently.
It has no physical significance.
For the $\DA$ method (\S\ref{bub}) 
the adjoint is simply the negative Hermitian of the direct matrix:
$\mathsfbi{\Phi}=-\mathsfbi{\Gamma}^H$.

\begin{figure}
\begin{center}
\includegraphics[width=0.99\textwidth, draft = false]{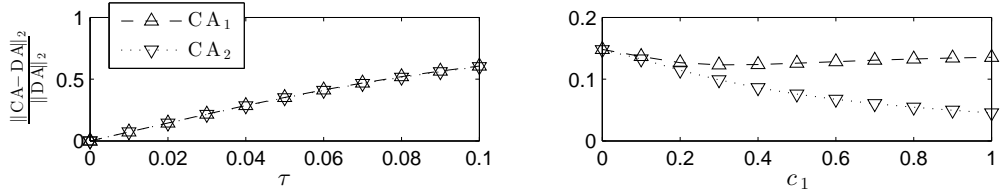}
\caption{The discrepancy between the discrete adjoint ($\DA$) and the continuous adjoint ($\CA$) discretizations, for the two different formulations of the continuous adjoint equations, $\CAone$ and $\CAtwo$. In both figures, $\mathit{N}=10$, $\mathit{x_h}=0.25$, and $\mathit{\beta}=0.5$. The left figure has $\mathit{c_1}=0.01$, $\mathit{c_2}=0.004$ and the right figure has $\mathit{c_2}=0$, $\mathit{\tau}=0.01$.}
\label{fig:1A2}
\end{center}
\end{figure}

The $\DA$ method has the same truncation errors
as the discretized direct system,
while methods $\CAone$ and $\CAtwo$ have different truncation errors.
The effect of these truncation errors is quantified in 
figure \ref{fig:1A2}, which compares
the discrepancy between $\CAone$ and $\DA$ 
with the discrepancy between $\CAtwo$ and $\DA$.
Method $\CAone$ has generally a greater discrepancy than $\CAtwo$, as shown in fig. \ref{fig:1A2}.
This discrepancy is a function of the time-delay, $\mathrm{\tau}$ and the damping coefficients, $\mathit{c_1}$ and $\mathit{c_2}$. Regardless of the value of the damping, the discrepancy is zero when $\mathrm{\tau}=0$.
%\textcolor{Red}{This is only a mathematical feature of this system.}
% I DON'T THINK THIS NEEDS TO BE SAID
This can be inferred by examining the mathematical structure of the matrices, given in eq. (\ref{gammaAPP})-(\ref{phi1}) and (\ref{phi2}). If $\mathrm{\tau}=0$ then $\mathsfbi{\Phi}=-\mathsfbi{\Gamma}^H$ regardless of the formulation used.

These adjoint equations govern the evolution of the adjoint variables, which can be regarded as Lagrange multipliers from a constrained optimization perspective \citep{BELEGUNDU-85}. 
Therefore, $u^+$ is the Lagrange multiplier of the acoustic momentum equation (\ref{equ_gov_mom_ndim}). Physically, it reveals the spatial distribution of the system's sensitivity to a force. Likewise, $p^+$ is the Lagrange multiplier of the pressure equation (\ref{equ_gov_enr_ndim}) \& (\ref{qqq1}) as well as (\ref{equ_gov_enr_ndim}) \& (\ref{qqq2}). Physically, it reveals the spatial distribution of the system's sensitivity to heat injection.
\begin{table}
\begin{center} 
  \begin{tabular}{  c | c } 
\hline
     $\CAone$ & $\CAtwo$ \\ \hline\hline
    \large{$\frac{\partial u^+}{\partial t} + \frac{\partial p^+}{\partial x}  +\beta\left(p^+ +\tau \frac{\partial p^+}{\partial t}\right)\delta(x-x_h)=0$}   & \large{$\frac{\partial u^+}{\partial t} + \frac{\partial p^+}{\partial x}  +\beta p^+\delta(x-x_h)=0$}   \\ 
     \large{$\frac{\partial u^+}{\partial x} + \frac{\partial p^+}{\partial t} - \zeta p^+=0$} & \large{$\frac{\partial u^+}{\partial x} + \frac{\partial p^+}{\partial t} - \zeta p^+-\beta\tau \frac{\partial [p^+\delta(x-x_h)]}{\partial x}=0$} \\ \hline
  \end{tabular}
\end{center}
\caption{The two different sets of continuous adjoint equations.}\label{tab_adj}
\end{table}
%%%%%%%%%%%%%%%%%%%%%%
% FIGURE%
%%%%%%%%%%%%%%%%%%%%%%
%
%
\section{Modal analysis: the eigenvalue problem} \label{normal}
So far 
we have considered the thermo-acoustic system in the ($\mathit{x}$, $\mathit{t}$) domain.
In modal analysis, 
we consider it in the ($\mathit{x}$, $\mathit{\sigma}$) domain using the transformations
\begin{equation}
u(x,t)=\hat{u}(x,\sigma)e^{\sigma t}, \;\;\;\;
u^{+}(x,t)=\hat{u}^{+}(x,\sigma)e^{-\bar{\sigma} t}, \label{no1}
\end{equation}
\begin{equation}
p(x,t)=\hat{p}(x,\sigma)e^{\sigma t}, \;\;\;\; p^{+}(x,t)=\hat{p}^{+}(x,\sigma)e^{-\bar{\sigma} t}.
\label{no3}
\end{equation}
where the symbol $\;\hat{ }\;$ denotes an eigenfunction. 
The behaviour of the system in the long time limit 
is dominated by the eigenfunction 
whose eigenvalue has the highest growth rate.
The complex conjugate adjoint eigenfunctions of velocity and pressure are labelled $\mathit{\hat{\bar{u}}}$ and $\mathit{\hat{\bar{p}}}$, respectively.
With the definition of the Green's identity (\ref{Greens_gen}),
the adjoint eigenvalues, $\mathit{-\bar{\sigma}}$,
are the negatives of the complex conjugates of the direct eigenvalues, $\mathit{\sigma}$.
This satisfies the bi-orthogonality condition
between the direct and adjoint eigenfunctions \citep{SALWEN-81}. 
The system is studied in the complex frequency domain by substituting the relations (\ref{no1})--(\ref{no3}) into the direct equations (\ref{equ_gov_mom_ndim})-(\ref{equ_gov_enr_ndim})-(\ref{qqq1}) and into the adjoint equations given in table \ref{tab_adj}.
\par
\begin{figure}
\begin{center}
\includegraphics[width=0.99\textwidth, draft = false]{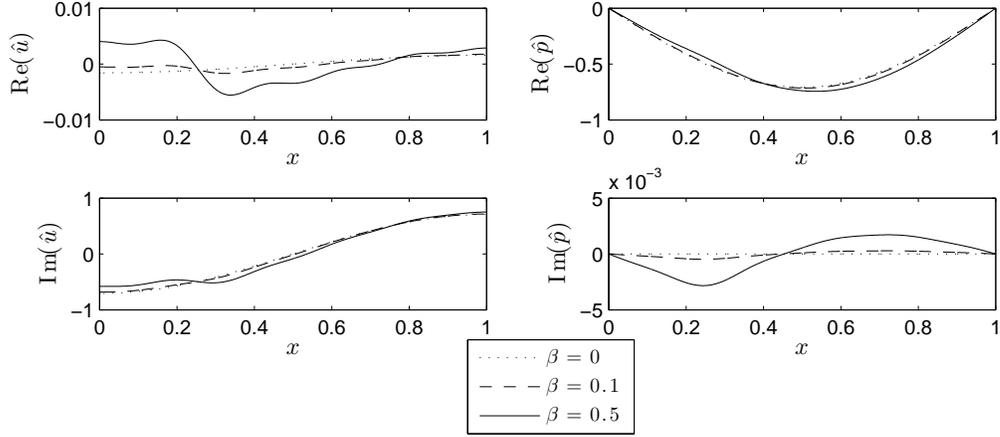}
\caption{The direct eigenfunctions as a function of the heat-release parameter, $\mathit{\beta}$, for $\mathit{N}=10$, $\mathit{x_h}=0.25$, $\mathit{\tau}=0.01$, $\mathit{c_1}=0.01$ and $\mathit{c_2}=0.004$. The relevant eigenvalues are: $\sigma=-0.0070+3.1416\mathrm{i}$, for $\beta=0$; $\sigma=-0.0056+3.1848\mathrm{i}$, for $\beta=0.1$; $\sigma=+0.00023+3.3570\mathrm{i}$, for $\beta=0.5$. Note that the top left and bottom right frames have very small vertical scales.} 
\label{fig:dE}
\end{center}
\end{figure}
\begin{figure}
\begin{center}
\includegraphics[width=0.99\textwidth, draft = false]{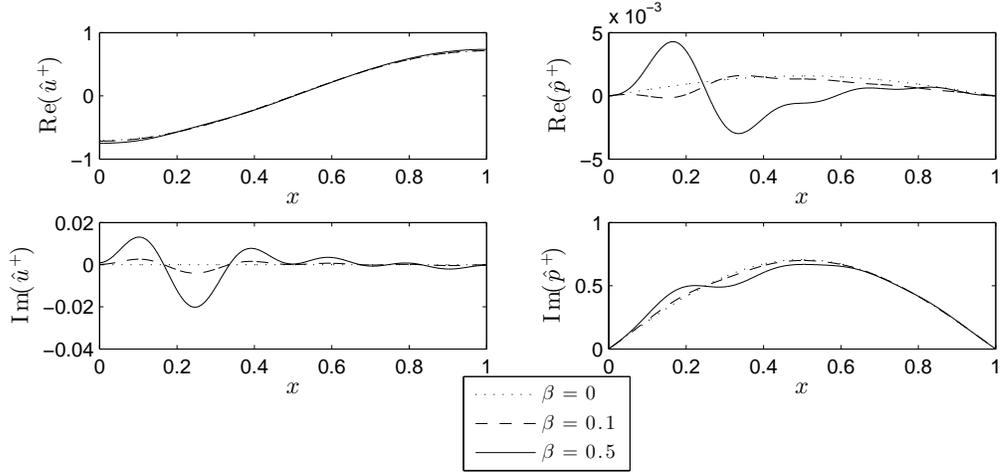}
\caption{The adjoint eigenfunctions as a function of the heat-release parameter, $\mathit{\beta}$, for $\mathit{N}=10$, $\mathit{x_h}=0.25$, $\mathit{\tau}=0.01$, $\mathit{c_1}=0.01$ and $\mathit{c_2}=0.004$. Note that the top right and bottom left frames have very small vertical scales.} 
\label{fig:aE}
\end{center}
\end{figure}

Figure \ref{fig:dE} shows the direct eigenfunctions and figure \ref{fig:aE} the $\DA$ adjoint eigenfunctions
%found using the $DA$ method, 
as $\mathit{\beta}$ increases from $0$ to $0.5$.
When $\mathit{\beta} = 0$, the eigenfunctions are the natural acoustic modes of the duct
but, as $\mathit{\beta}$ increases, the eigenfunctions become distorted by the heat release at the wire.
This has important consequences for the structural sensitivity,
as will be shown in \S \ref{rpi}.
\par
\begin{figure}
\begin{center}
\includegraphics[width=0.99\textwidth, draft = false]{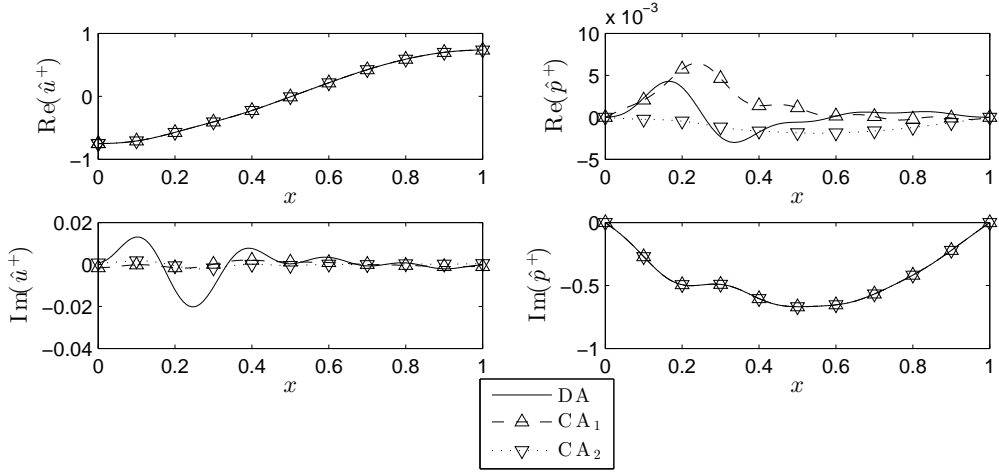}
\caption{The adjoint eigenfunctions found using the $\DA$, $\CAone$, and $\CAtwo$ methods. The parameters are $\mathit{N}=10$, $\mathit{x_h}=0.25$, $\mathit{\tau}=0.01$, $\mathit{\beta}=0.5$, $\mathit{c_1}=0.01$ and $\mathit{c_2}=0.004$. Note that the top right and bottom left frames have very small vertical scales.} 
\label{fig:DAE}
\end{center}
\end{figure}

Figure \ref{fig:DAE} shows the direct and adjoint eigenfunctions,
found using the $\DA$, $\CAone$, and $\CAtwo$ methods,
at $\mathit{\beta} = 0.5$.
This is the value of $\mathit{\beta}$ used for the sensitivity analyses.
The  discrepancies in $\Imag\mathit{(u^+)}$ and $\Real\mathit{(p^+)}$ cause the differences in sensitivities seen in \S \ref{rpi_tm} and \S\ref{bfsr}.
%
%%%%%%%%%%%%%%%%%%%%%%
% FIGURE%
%%%%%%%%%%%%%%%%%%%%%%
%%%%%%%%%%%%%%%%%%%%%%
% FIGURE%
%%%%%%%%%%%%%%%%%%%%%%
%%%%%%%%%%%%%%%%%%%%%%
% FIGURE%
%%%%%%%%%%%%%%%%%%%%%%
%%%%%%%%%%%%%%%%%%%%%%
%
%%%%%%%%%%%%%%%%%%%%%%%%%%%%%%%%%%%%%%%%%%%%%%%%%%%%%%%%%%%%%%%%%%%%%%%%%%
\section{Calculation of the structural and base-state sensitivities} \label{csb}
\subsection{Structural sensitivity via the $\CA$ method}\label{tw}
The thermo-acoustic system described in \S \ref{sec_mod_def} has been linearized about a base state. 
Following \citet{GIANNETTI-07}, 
we perturb the linearized operator, $\mathrm{L}$, 
by adding to it
some general function of the perturbation state variables, $\mathit{\hat{u}}$ and $\mathit{\hat{p}}$. 
In this section, 
we assume that this feedback
%between the perturbation variables and the linearized operator 
does not affect the base state. 
We also assume that the structural perturbation 
is small enough for the new thermo-acoustic configuration to be
\begin{equation}
\sigma_{new}=\sigma+\delta\sigma, \;\;\;\; \hat{p}_{new}=\hat{p}+\delta \hat{p}, \;\;\;\; \hat{u}_{new}=\hat{u}+\delta\hat{u},     \label{decos}
\end{equation}
where $\mathit{\delta\sigma=\epsilon\sigma}$, $\mathit{\delta \hat{p}=\epsilon\hat{p}}$, $\mathit{\delta\hat{u}=\epsilon\hat{u}}$ with $|\mathit{\epsilon}|\ll 1$,
and where terms of order $\epsilon^2$ are sufficiently small to be neglected.
\par
The direct eigenfunctions can be arranged as a column vector $\mathit{[\hat{u}\;\;\hat{p}]^T}$.
In general, a structural perturbation to the thermo-acoustic operator can be represented by a $2\times2$ tensor, $\mathit{\delta} \mathsfbi{H}$, which acts on $\mathit{[\hat{u}\;\;\hat{p}]^T}$.
Each component $\mathit{\delta} \mathsfbi{H_{ij}}$ of this structural perturbation tensor 
quantifies the effect of a feedback mechanism between the $\mathit{j^{th}}$ eigenfunction 
and the $\mathit{i^{th}}$ direct governing equation.
\par
We obtain the eigenvalue drift, $\mathit{\delta \sigma}$, 
caused by the structural perturbation, $\mathit{\delta} \mathsfbi{H}$, %(table \ref{dS}) 
by applying the Green\rq{}s identity (\ref{Greens_gen}) 
to the perturbed direct and adjoint equations, 
in a manner similar to \citet{GIANNETTI-07}. 
Table \ref{dS} describes the effect of a generic perturbation $\mathbf{\delta} \mathsfbi{H}$.
The great advantage of this approach is that,
once the direct and adjoint eigenfunctions have been calculated, 
all linear feedback mechanisms can be examined at little extra cost. 
\par
We will illustrate the process for the specific case where
the feedback mechanism is a second hot wire, called the control wire,
whose parameters are denoted with the subscript $c$.
The structural perturbation caused by the control wire is represented by the tensor in table \ref{deltaH}.
\begin{table} 
\begin{center}
  \begin{tabular}{ l || c | c}
\hline
    Method & $\CAone$ & $\CAtwo$ \\ \hline\hline
    $\delta\sigma=$ & \Large{$\frac{\int_L \! [\hat{\bar{u}}^+\;\; \hat{\bar{p}}^+]\delta\mathsfbi{H}[\hat{u}\;\; \hat{p}]^T \mathrm{d} x}{\int_L \! (\hat{u}\hat{\bar{u}}^+  + \hat{p}\hat{\bar{p}}^+) \mathrm{d} x+\beta\tau\hat{u}_h \hat{\bar{p}}^+_h} \label{drift_rik}$} & \Large{$\frac{\int_L \! [\hat{\bar{u}}^+ \;\;\hat{\bar{p}}^+]\delta\mathsfbi{H}[\hat{u} \;\;\hat{p}]^T \mathrm{d} x}{\int_L \! (\hat{u}\hat{\bar{u}}^+  + \hat{p}\hat{\bar{p}}^+) \mathrm{d} x} \label{drift_rik2}$} \\ \hline
  \end{tabular}
\end{center}
\caption{The eigenvalue drift caused by a generic structural perturbation, which is represented by the generic tensor $\mathit{\delta}\mathsfbi{H}$. The two methods, $\CAone$ and $\CAtwo$, are derived from two equivalent versions of the governing equations, \S \ref{adjoints}. $\mathit{L}$ is the dimensionless tube length.}\label{dS}
\end{table}
\begin{table}
\begin{center}
  \begin{tabular}{ l || c | c}
\hline
    Method & $\CAone$ & $\CAtwo$ \\ \hline\hline
    $\delta\mathsfbi{H}=$ & $\begin{bmatrix} 0 & 0 \\ \delta\beta_c (1-\sigma\tau_c)\delta(x-x_c) &0 \end{bmatrix}$ & $\begin{bmatrix} 0 & 0 \\ \delta\beta_c \delta(x-x_c) & \delta\beta_c\tau_c\delta(x-x_c)\frac{\partial}{\partial x} \end{bmatrix}$ \\ \hline
  \end{tabular}
\end{center}
\caption{The tensor representing a structural perturbation caused by a second hot wire. Two representations are obtained, depending on whether the heat-release rate is expressed following the $\CAone$ or $\CAtwo$ method.}\label{deltaH}
\end{table}
The component $\mathit{\delta}\mathsfbi{H_{21}}$ represents 
a feedback mechanism
that is proportional to the velocity perturbation
and that perturbs the pressure equation. 
The component $\mathit{\delta} \mathsfbi{H_{22}}$ represents 
a feedback mechanism
that is proportional to the pressure perturbation
and that also perturbs the pressure equation.
The change in the eigenvalue caused by the presence of the control hot wire with a small heat-release parameter $\mathit{\delta\beta_c}$ is given in table \ref{tab_Struc} for both $\CA$ methods. The results will be described in section \ref{cssri}. 
\begin{table}
\begin{center}
  \begin{tabular}{ l || c | c}
\hline
    Method & $\CAone$ & $\CAtwo$ \\ \hline\hline
    \Large{$\frac{\delta\sigma}{\delta\beta_c}=$} & \Large{$\frac{ \hat{\bar{p}}^+_c\hat{u}_c\left(1-\sigma\tau_c\right)}{\int_L \! (\hat{u}\hat{\bar{u}}^+  + \hat{p}\hat{\bar{p}}^+) \mathrm{d} x+\beta\tau \hat{u}_h \hat{\bar{p}}^+_h}$} & \Large{$\frac{\hat{\bar{p}}^+_c\left(\hat{u}_c+\tau_c\frac{\partial \hat{p}_c}{\partial x}\right)}{\int_L \! (\hat{u}\hat{\bar{u}}^+  + \hat{p}\hat{\bar{p}}^+) \mathrm{d} x}$} \\ \hline
  \end{tabular}
\end{center}
\caption{The change in the eigenvalue due to the presence of the control wire with a small heat-release parameter $\mathit{\delta\beta_c}$, derived via the $\CAone$ and $\CAtwo$ approaches.}\label{tab_Struc}
\end{table}
%
%
%%%%%%%%%%%%%%%%%%%%%%%%%%%%%%%%%%%%%%%%%%%%%%%%%%%%%%%%%%%%%%%%%%%%%%%%%%%%%%%
\subsection{Base-state sensitivity via the CA method} \label{fff}
Using adjoint techniques, 
a single calculation can reveal
how the growth rate and frequency of the system is altered 
by any small variation of the base-state parameters $\mathit{\delta\beta}$, $\mathit{\delta\zeta}$, $\mathit{\delta\tau}$, and $\mathit{\delta x_h}$. This is known as the base-state sensitivity. 
In this section, we calculate the base-state sensitivities to $\mathit{\beta}$, $\mathit{\tau}$, and $\mathit{\zeta}$ as functions of the hot wire position, $\mathit{x_h}$.
By applying a methodology similar to that presented in \S \ref{tw}, 
we obtain the base-state sensitivities shown in table \ref{tab_base}.
\begin{table}
\begin{center}
  \begin{tabular}{ l || c | c }
\hline
    Method & $\CAone$ & $\CAtwo$ \\ \hline\hline
    \Large{$\frac{\delta\sigma}{\delta\beta}=$} & \Large{$\frac{ \hat{\bar{p}}^+_h\hat{u}_h\left(1-\sigma\tau\right)}{\int_L \! (\hat{u}\hat{\bar{u}}^+  + \hat{p}\hat{\bar{p}}^+) \mathrm{d} x+\beta\tau \hat{u}_h \hat{\bar{p}}^+_h} \label{drift_beta}$} & \Large{$\frac{ \hat{\bar{p}}^+_h\left(\hat{u}_h+\tau\frac{\partial \hat{p}_h}{\partial x}\right)}{\int_L \! (\hat{u}\hat{\bar{u}}^+  + \hat{p}\hat{\bar{p}}^+) \mathrm{d} x } \label{drift_beta2}$} \\ \hline
 \Large{$\frac{\delta\sigma}{\delta\tau}=$} & \Large{$\frac{-\beta\sigma\hat{\bar{p}}^+_h\hat{u}_h}{\int_L \! (\hat{u}\hat{\bar{u}}^+  + \hat{p}\hat{\bar{p}}^+) \mathrm{d} x+\beta\tau \hat{u}_h \hat{\bar{p}}^+_h} \label{drift_tau}$} & \Large{$\frac{\beta \hat{\bar{p}}^+_h\frac{\partial \hat{p}_h}{\partial x}}{\int_L \! (\hat{u}\hat{\bar{u}}^+  + \hat{p}\hat{\bar{p}}^+) \mathrm{d} x} \label{drift_tau2}$} \\ \hline
 \Large{$\frac{\delta\sigma}{\delta\zeta}=$} & \Large{$\frac{-\int_L \! \hat{p}\hat{\bar{p}}^+ \mathrm{d} x}{\int_L \! (\hat{u}\hat{\bar{u}}^+  + \hat{p}\hat{\bar{p}}^+) \mathrm{d} x+\beta\tau \hat{u}_h\hat{\bar{p}}^+_h} \label{drift_zeta}$} & \Large{$\frac{-\int_L \! \hat{p}\hat{\bar{p}}^+ \mathrm{d} x}{\int_L \! (\hat{u}\hat{\bar{u}}^+  + \hat{p}\hat{\bar{p}}^+) \mathrm{d} x} \label{drift_zeta2}$} \\ \hline
  \end{tabular}
\end{center}
\caption{The change in the eigenvalue due to small changes in the base-state coefficients, derived via the $\CAone$ and $\CAtwo$ methods.}\label{tab_base}
\end{table}
%
%As an aside,
%the damping factor for each Galerkin mode varies according to
%$\delta\zeta_j=\delta c_{1} j^2 + \delta c_{2}\sqrt{j}$.
%This means that
%the sensitivity coefficient with respect to $\delta\zeta$ in table \ref{tab_base} 
%must be interpreted as 
%$\frac{\delta\sigma}{\delta\zeta}=\sum_{j=1}^N \frac{\delta\sigma}{\delta\zeta_j}\frac{\delta\zeta_j}{\delta\zeta}$, where $\delta\zeta=\sum_{j=1}^N {\delta\zeta_j}$.
%The sensitivity
%$\frac{\delta\sigma}{\delta\zeta_j}$ is calculated by using only the $j^{th}$ eigenmodes in the formula provided in the bottom row of table \ref{tab_base}. 
%
The results will be described in \S \ref{bfsr}.
%
%
%%%%%%%%%%%%%%%%%%%%%%%%%%%%%%%%%%%%%%%%%%%%%%%%%%%%%%%%%%%%%%%%%%%
%
\subsection{Both sensitivities via the DA method}\label{bub}
Both sensitivities can be calculated directly from the discretized governing equations (the $\DA$ method). 
There are four stages to this method: 
(1) calculate the perturbation matrix $\mathit{\delta}\mathsfbi{P}$ using (\ref{we}), imposing an arbitrarily small perturbation on the base-state parameter; 
(2) calculate the eigenvectors of the matrices $\mathsfbi{\Gamma}$ and $-\mathsfbi{\Gamma}^H$; 
(3) apply (\ref{gin}) to find the eigenvalue drift; 
(4) divide the eigenvalue drift by the perturbation used in stage 1 in order to obtain the sensitivity coefficient. 
The eigenvalue drift due to a perturbation of the discretized direct system (similar to \citet{GIANNETTI-07}) is given by
\begin{equation}
 \delta \sigma=\frac{\boldsymbol{\hat{\bar{\xi}}} \cdot \left( \delta \mathsfbi{P}\hat{\boldsymbol{\chi}}\right)}{\boldsymbol{\hat{\bar{\xi}}} \cdot \boldsymbol{\hat{\chi}}} . \label{gin}
\end{equation}
where the matrix $\mathit{\delta}\mathsfbi{P}$ represents 
a small perturbation to the direct system, 
whose matrix is $\mathsfbi{\Gamma}$.  
Here, the symbol $\;\hat{ }\;$ represents an eigenvector. 
The column vector $\boldsymbol{\hat{\xi}}$ is the eigenvector of the adjoint matrix 
$\mathsfbi{\Phi}=-\mathsfbi{\Gamma}^H$.
The perturbation matrix $\mathit{\delta}\mathsfbi{P}$ is described in (\ref{we}). 
It can represent either a structural perturbation or a base-state perturbation. 
%
%
%%%%%%%%%%%%%%%%%%%%%%%%%%%%%%%%%%%%%%%%%%%%%%%%%%%%%%%%%%%%%%%%%
\section{Results and physical interpretation} \label{rpi}
\subsection{Comparing the three methods of calculating the structural sensitivity} \label{rpi_tm}
\par
\begin{figure}
\begin{center}
\includegraphics[width=0.99\textwidth, draft = false]{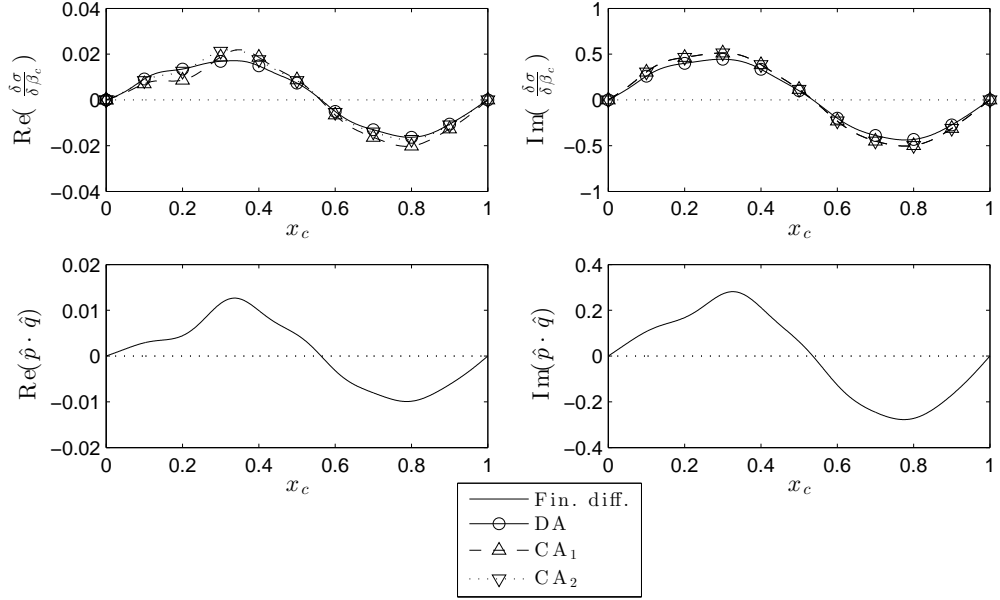}
\caption{Top frames: sensitivity of the growth rate, \Real($\mathit{\delta\sigma/\delta\beta_c}$), and of the angular frequency, \Imag($\mathit{\delta\sigma/\delta\beta_c}$), when a control wire is placed at position $\mathit{x_c}$. This is calculated exactly, via finite difference, and via the $\DA$, $\CAone$ and $\CAtwo$ methods. (The $\DA$ method gives the same result as the finite difference method to machine precision.) Bottom frames: the Rayleigh index for a control wire placed at $\mathit{x_c}$. The parameters are $\mathit{N}=10$, $\mathit{\beta}=0.5$, $\mathit{c_1}=0.01$, $\mathit{c_2}=0.004$ and $\mathit{\tau=\tau_c}=0.01$. The main hot wire is at $\mathit{x_h}=0.25$ so that the first acoustic mode is excited.} 
\label{fig:2A2}
\end{center}
\end{figure}
\begin{figure}
\begin{center}
\includegraphics[width=0.99\textwidth, draft = false]{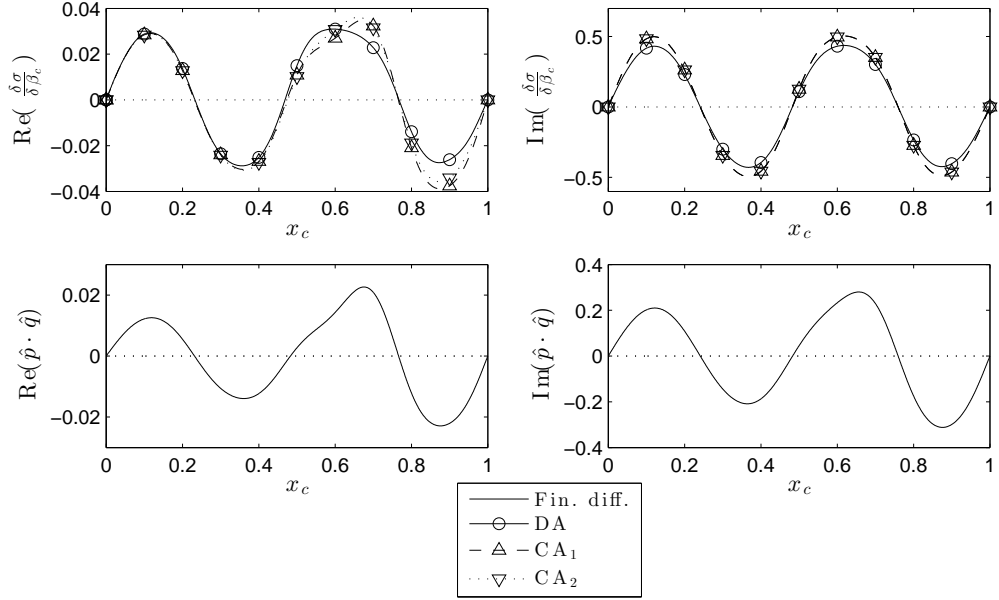}
\caption{Top frames: sensitivity of the growth rate, \Real($\mathit{\delta\sigma/\delta\beta_c}$), and of the angular frequency, \Imag($\mathit{\delta\sigma/\delta\beta_c}$), when a control wire is placed at position $\mathit{x_c}$. This is calculated exactly, via finite difference, and via the $\DA$, $\CAone$ and $\CAtwo$ methods. (The $\DA$ method gives the same result as the finite difference method to machine precision.) Bottom frames: the Rayleigh index for a control wire placed at $\mathit{x_c}$. The parameters are $\mathit{N}=10$, $\mathit{\beta}=0.5$, $\mathit{c_1}=0.01$, $\mathit{c_2}=0.004$ and $\mathit{\tau=\tau_c}=0.01$. The main hot wire is at $\mathit{x_h}=0.625$ so that the second acoustic mode is excited.} 
\label{fig:2A22}
\end{center}
\end{figure}

The top frames of figure \ref{fig:2A2} show the real and imaginary components of $\mathit{\delta \sigma / \delta \beta_c}$ as a function of the control wire position, $\mathit{x_c}$, via the $\DA$, $\CAone$ and $\CAtwo$ methods. In this case the main hot wire is placed at $x=0.25$ so that most of the perturbation energy is in the first acoustic mode \citep{MATVEEV-03}.
Similarly, the top frames of figure \ref{fig:2A22} show the real and imaginary components of $\mathit{\delta \sigma / \delta \beta_c}$ as a function of the control wire position, $\mathit{x_c}$, via the $\DA$, $\CAone$ and $\CAtwo$ methods. In this case the main hot wire is placed at $x=0.625$ so that most of the perturbation energy is in the second acoustic mode \citep{MATVEEV-03}.
These results can be compared with the exact solution,
which is obtained by finite difference.
This is the difference between 
the dominant eigenvalues of the perturbed direct matrix,
$\mathsfbi{\Gamma}+\mathit{\delta}\mathsfbi{P}$,
and the original direct matrix, 
$\mathsfbi{\Gamma}$, divided by the (finite) arbitrarily small perturbation.
The perturbation matrix $\mathit{\delta}\mathsfbi{P}$ is given in (\ref{we}).  
\par
As expected, the $\DA$ method matches the finite difference method exactly. 
The $\CA$ methods both contain some error, due to the truncation errors in the discretization. %, but these do not exceed 10\%. %, in agreement with \cite{CHANDLER-12}.
%
%The discrepancy is high in  Re($\delta\sigma/\delta\beta_c$) at $x\approx0.9$, as depicted in the top left frame of figure \ref{fig:2A22}. 
% THERE IS NO NEED TO POINT THIS OUT - THE READER CAN SEE FOR THEMSELVES
The $\CAtwo$ method is usually more accurate than the $\CAone$ method.
For this thermo-acoustic system, however, the $\DA$ method turns out to be the most accurate and easy to implement. 
\par
The real component of the structural sensitivity gives the change in the growth rate that is caused by the control wire. 
The imaginary component gives the change in the frequency. 
The physical reason for these changes is given in \S \ref{cssri}.
The control wire has a much stronger effect on the frequency than on the growth rate, for reasons given in \S \ref{fefb}.
%
%%%%%%%%%%%%%%%%%%%%%%
% FIGURE%
%%%%%%%%%%%%%%%%%%%%%%
\subsection{Comparing the structural sensitivity with the Rayleigh Index} \label{cssri}
\par
It has long been known \citep{RAYLEIGH-1878} that if pressure and heat-release fluctuations are in phase, then acoustic vibrations are encouraged. 
More precisely, the Rayleigh criterion states that the energy of the acoustic field grows over one cycle of oscillation if
$
\oint_T \int_{\mathcal{D}}p \dot{q} \;\mathrm{d}\mathcal{D} \mathrm{d} t \; ,
$
exceeds the damping,
%\end{equation}
%
where $\mathcal{D}$ is the flow domain and $\mathit{T}$ is the period.
It is particularly informative to plot the spatial distribution of 
\begin{equation} \label{Rayl_ind}
\oint_T p \dot{q} \;\mathrm{d} t
\end{equation}
which is known as the Rayleigh Index.
This reveals the regions of the flow that contribute most to the Rayleigh Criterion
and therefore gives insight into the physical mechanisms that alter the amplitude of the oscillation. 
To examine the effect of the control wire, 
we substitute the approximate expressions
$\mathit{p} = \mathit{\hat{p} \exp(\sigma_i t)}$ and 
$\mathit{\dot{q}} = \mathit{\hat{\dot{q}} \exp(\sigma_i t)}$ (found from \ref{qqq1} or \ref{qqq2})
into (\ref{Rayl_ind})
and integrate over a period $2\upi / \mathit{\sigma_i}$,
where $\mathit{\sigma_i} = \Imag\mathit{(\sigma)}$.
(The approximation arises because the growth rate over the cycle has been ignored.)
The real part of the Rayleigh Index gives the change in the growth rate and the imaginary part gives the change in the frequency (figure \ref{fig:2A2}-\ref{fig:2A22}). 
As expected, the sign of the Rayleigh index  matches that of the structural sensitivity (the position at which it is zero matches within 1\%) and the shape is similar. 
The Rayleigh Index physically explains the effect of adding the control hot wire to the Rijke tube.
\\ \\
Firstly, we refer to fig.~\ref{fig:2A2} where the main hot wire is at $\mathit{x_h}=0.25$ and most of the perturbation energy is contained in the first mode.
For $\mathit{x} = 0$ to $0.56$, the pressure and heat release eigenfunctions are sufficiently in phase that the contribution to growth over a cycle is positive. For $\mathit{x} = 0.56$ to $1$, they are out of phase so their contribution to growth over a cycle is negative. 
For this case, the  location where the presence of a second hot wire is most effective at reducing the growth rate is $x_c\approx0.8$.
%
%The most stabilising location is $x_c=0.81$ and the most destabilising is $x_c=0.32$. 
%
It is interesting to note that this system becomes more unstable 
when the control wire is placed at $0.5 <\mathit{x_c} < 0.56$.
This is in the second half of the tube
and, in the absence of the first hot wire, 
a control wire placed here would be stabilizing.
The reason for this is that 
the main hot wire, at $\mathit{x_h}$, 
causes the eigenfunctions to distort from the acoustic modes of the duct.
In particular, the features of the $\mathit{\hat{u}}$ and $\mathit{\hat{p}}$ eigenfunctions (figure \ref{fig:dE}) 
shift down the duct, to higher values of $\mathit{x}$.
This shifts downstream the region in which the control wire is destabilizing.
\\ \\
Secondly, we refer to figure \ref{fig:2A22} where the main hot wire is at $\mathit{x_h}=0.625$ and most of the perturbation energy is contained in the second mode.
For $0<\mathit{x} <0.23$ and $0.47<\mathit{x} <0.77$, the pressure and heat release eigenfunctions are sufficiently in phase that the contribution to growth over a cycle is positive; 
for $0.23<\mathit{x} <0.47$ and $0.77<\mathit{x} <1$, they are out of phase so their contribution to growth over a cycle is negative. 
For this case, the  location where the presence of a second hot wire is most effective at reducing the growth rate is $x_c\approx0.36$. 
%
%The most stabilising location is $x_c=0.81$ and the most destabilising is $x_c=0.32$. 
%
%
%
\subsection{Using the structural sensitivity to find the most efficient feedback mechanisms} \label{fefb}
\par
In passive control,
an object that is placed at a point in the system
causes feedback at that point.
Under these conditions,
the structural sensitivity 
reveals the feedback mechanism 
that is most effective at changing
the frequency or growth rate of the system.
%In other words, the sensor and actuator are co-located. 
%
\par
\begin{figure}
\begin{center}
\includegraphics[width=0.99\textwidth, draft = false]{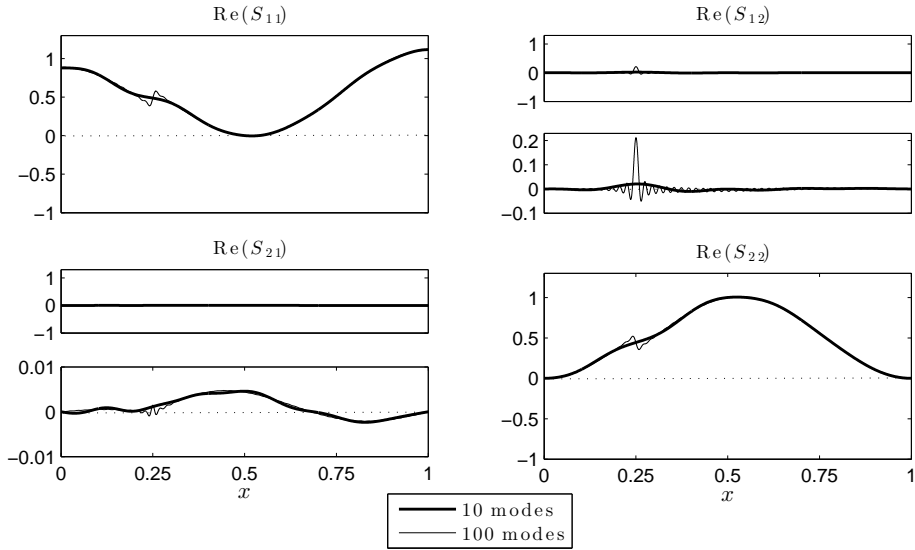}
\caption{Real part of the components of the structural sensitivity tensor (via the $\DA$ method) for the same parameters as fig. \ref{fig:2A2}. These components indicate the effect of a feedback mechanism on the growth rate of oscillation.} 
\label{fig:str1}
\end{center}
\end{figure}
\begin{figure}
\begin{center}
\includegraphics[width=0.99\textwidth, draft = false]{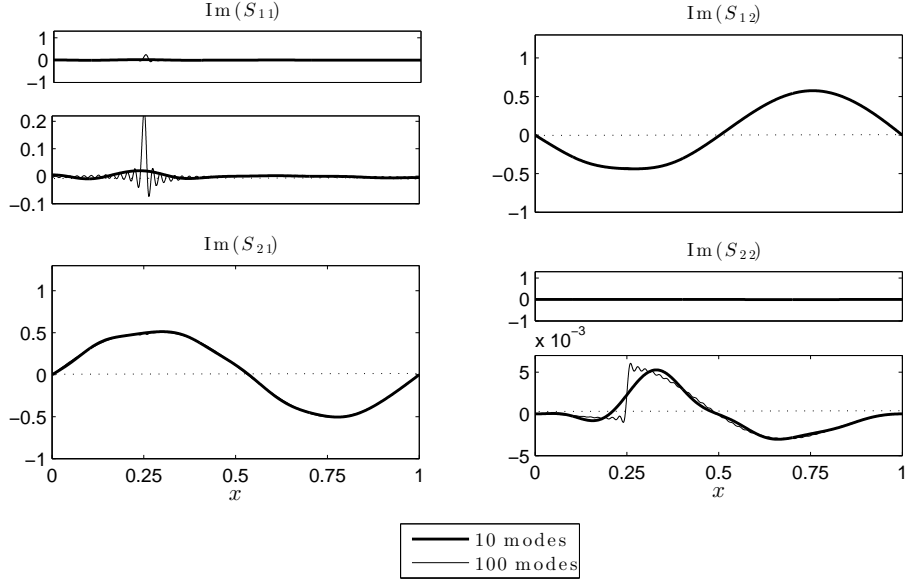}
\caption{Imaginary part of the components of the structural sensitivity tensor (via the $\DA$ method) for the same parameters as fig. \ref{fig:2A2}. These components indicate the effect of a feedback mechanism on the angular frequency of oscillation.} 
\label{fig:str1B}
\end{center}
\end{figure}

%The perturbation tensor $\delta\mathbf{H}$ is, in general, an operator function of the space variable. \textcolor{Red}{I understand this sentence now, but I think it could be clearer - not sure how yet though.}
%
In \S \ref{tw}, we defined the perturbation tensor, $\mathit{\delta}\mathsfbi{H}$,
to be an operator localized at the control wire's location.
%, by using a delta function. 
%
%Now we will examine the general case in which 
%the forcing is a linear function of all the state variables at that point. \textcolor{Red}{$<$---It's a little misleading, can we remove it?}
%,
%which is always true in the limit of small forcing.  \textcolor{Red}{---I 'd cut this sentence off because is a bit redundant. In linear analysis it always holds true}
%\textcolor{Red}{This is always the case for small feedback, isn't it?}.
%
In this section we consider the case of a generic feedback mechanism,
represented by a localized perturbation in which $\mathit{\delta}\mathsfbi{H}$ is constant,
following \citet{GIANNETTI-07}.
For clarity, we re-label $\mathit{\delta}\mathsfbi{H}$ as $\mathit{\delta}\mathsfbi{C}$ for this case.  
The \emph{structural sensitivity tensor} 
$\mathsfbi{S}=\mathit{\delta\sigma} / \mathit{\delta}\mathsfbi{C}$ 
is then given by the expression in table \ref{lasTAB}. 
Its numerator is the dyadic product $\mathit{[\hat{\bar{u}}^+\;\; \hat{\bar{p}}^+]^T}\otimes\mathit{[\hat{u}\;\; \hat{p}]^T}$.
The four components of $\mathsfbi{S}$ quantify how a feedback mechanism that is proportional to the state variables affects the growth rate and frequency of the system.
They are shown in fig.~\ref{fig:str1} (real part) and fig.~\ref{fig:str1B} (imaginary part) as a function of $\mathit{x}$, 
which is the location of the structural perturbation.
The eigenfunctions are calculated with both 10 modes (thick line) and 100 modes (thin line). 
With the latter discretization it is possible to capture the eigenfunction discontinuity at the hot wire's location caused by the impulsive heat release. 
Although a discretization with 100 modes does not meet the physical constraint that $\tau\ll 2/N$ (\S \ref{numdisc}), we can use it to examine the numerical accuracy of the 10-mode discretization. 
At the hot wire's location, the 100-mode discretization of Re$(\mathsfbi{S_{12}}$) and Im$(\mathsfbi{S_{11}}$) experiences the Gibbs phenomenon \citep{GIBBS-1898} therefore the solution is inaccurate. The Gibbs phenomenon remains as the number of Galerkin modes increases. 
The 10-mode discretization is very accurate except at the discontinuity at the hot wire's location. 

When $\mathit{\beta} \rightarrow 0$, 
the direct eigenfunctions are nearly the acoustic modes of the system, as shown in fig. \ref{fig:dE}-\ref{fig:aE}.
By inspection of these eigenfunctions, we see that 
$\mathsfbi{S_{11}} \approx (\cos \upi x)^2$,
$\mathsfbi{S_{12}} \approx -\mbox{i} (\sin \upi x) \times (\cos \upi x)$,
$\mathsfbi{S_{21}} \approx \mbox{i} (\sin \upi x) \times (\cos \upi x)$, and
$\mathsfbi{S_{22}} \approx (\sin \upi x)^2$, when $\mathit{\beta} \rightarrow 0$.
The heat release from the main hot wire distorts these eigenfunctions 
(figure \ref{fig:dE})
so the structural sensitivities are similarly distorted.
\par
Firstly, we consider a feedback mechanism that is proportional to the velocity and that forces the momentum equation ($\mathsfbi{S_{11}}$). For example, this could be caused by the drag from a fine mesh placed in the flow. The system is most sensitive when this feedback mechanism is placed at the entrance or exit of the duct.
This is because (i) the velocity mode is maximal there and (ii) the adjoint velocity, which is a measure of the sensitivity of the momentum equation, is also maximal there,
as shown in figure \ref{fig:DAE}.
The Re($\mathsfbi{S_{11}}$) component (fig.~\ref{fig:str1}) is positive for all values of $\mathit{x}$, 
which means that, whatever value of $\mathit{x}$ is chosen,
the growth rate will decrease
if the forcing is in the opposite direction to the velocity,
as it would be for a fine mesh placed in the flow. 
This type of feedback greatly affects the growth rate (fig.~\ref{fig:str1}), which is the real component of the sensitivity, but barely affects the frequency (fig.~\ref{fig:str1B}), which is the imaginary component. 
This behaviour is as expected for this type of feedback.
%(This is as expected physically.)
%
\par
Secondly, we consider a feedback mechanism that is proportional to the pressure and that forces the pressure equation ($\mathsfbi{S_{22}}$). 
This type of feedback is described in \citet{CHU-63} and is relevant to pressure-coupled heat release in solid rocket engines.
For this feedback, the system is most sensitive around the centre of the duct, 
with a maximum at $\mathit{x}\approx0.58$.
Again, this feedback greatly affects the growth rate (fig.~\ref{fig:str1}), and it is positive for all values of $\mathit{x}$, but barely affects the frequency (fig.~\ref{fig:str1B}).
If the heat release increases with the pressure,
as it does for most chemical reactions,
this feedback mechanism is destabilizing. 
\par
Thirdly, 
we consider $\mathsfbi{S_{12}}$, 
which represents feedback from the pressure into the momentum equation
and $\mathsfbi{S_{21}}$, 
which represents feedback from the velocity into the pressure equation.
These types of feedback barely affect the growth rate (fig.~\ref{fig:str1}) but greatly affect the frequency (fig.~\ref{fig:str1B}). 
The hot control wire considered in figure \ref{fig:2A2} causes this type of feedback ($\mathsfbi{S_{21}}$) if $\tau=0$.
This analysis shows, therefore, that this passive control device is quite ineffective
at reducing the growth rate.
This had been seen already in figure \ref{fig:2A2}, in which the hot wire is seen to affect the frequency (imaginary component) much more than it affects the growth rate (real component). 
\par
By inspection of fig. \ref{fig:str1}, 
we conclude that the passive device 
that is most effective at reducing the growth rate
should force the momentum equation in the opposite direction to the velocity fluctuation
and should be placed at the exit of the tube. 
A damping device such as an adiabatic wire mesh would achieve this. 
\begin{table} 
\begin{center}
  \begin{tabular}{ l || c | c}
\hline
    Method & $\CAone$ & $\CAtwo$ \\ \hline\hline
    $\mathsfbi{S}=\large{\frac{\delta\sigma}{\delta\mathsfbi{C}}}=$ & \Large{$\frac{[\hat{\bar{u}}^+\;\; \hat{\bar{p}}^+]^T\otimes[\hat{u}\;\; \hat{p}]^T}{\int_L \! (\hat{u}\hat{\bar{u}}^+  + \hat{p}\hat{\bar{p}}^+) \mathrm{d} x+\beta\tau\hat{u}_h \hat{\bar{p}}^+_h} \label{drift_rik}$} & \Large{$\frac{[\hat{\bar{u}}^+ \;\;\hat{\bar{p}}^+]^T\otimes[\hat{u} \;\;\hat{p}]^T}{\int_L \! (\hat{u}\hat{\bar{u}}^+  + \hat{p}\hat{\bar{p}}^+) \mathrm{d} x} \label{drift_rik2}$} \\ \hline
  \end{tabular}
\end{center}
\caption{Structural sensitivity tensor for a general feedback mechanism $\delta\mathsfbi{C}$.}\label{lasTAB}
\end{table}

%%%%%%%%%%%%%%%%%%%%%%
% FIGURE%
%%%%%%%%%%%%%%%%%%%%%%
%
\par
This paper is mainly relevant to passive control 
but it is worth briefly mentioning active control.
For active control, 
the sensor and actuator would typically be in different places. For maximum observability, the sensor should be placed where the relevant direct eigenfunction has its largest amplitude. For maximum controllability, the actuator should be placed where the relevant adjoint eigenfunction has its largest amplitude. 
%For example, if the actuator forces the momentum equation, 
%it should be placed at the entrance or exit of the tube (figure \ref{fig:str1}-\ref{fig:dE}). %$x = 0.0$ (?). 
%
%The structural sensitivity is most relevant to passive control, however,
%so we will not consider active control further in this paper. 
%
\subsection{Base-state sensitivity results}\label{bfsr}
\par
\begin{figure}
\begin{center}
\includegraphics[width=0.99\textwidth, draft = false]{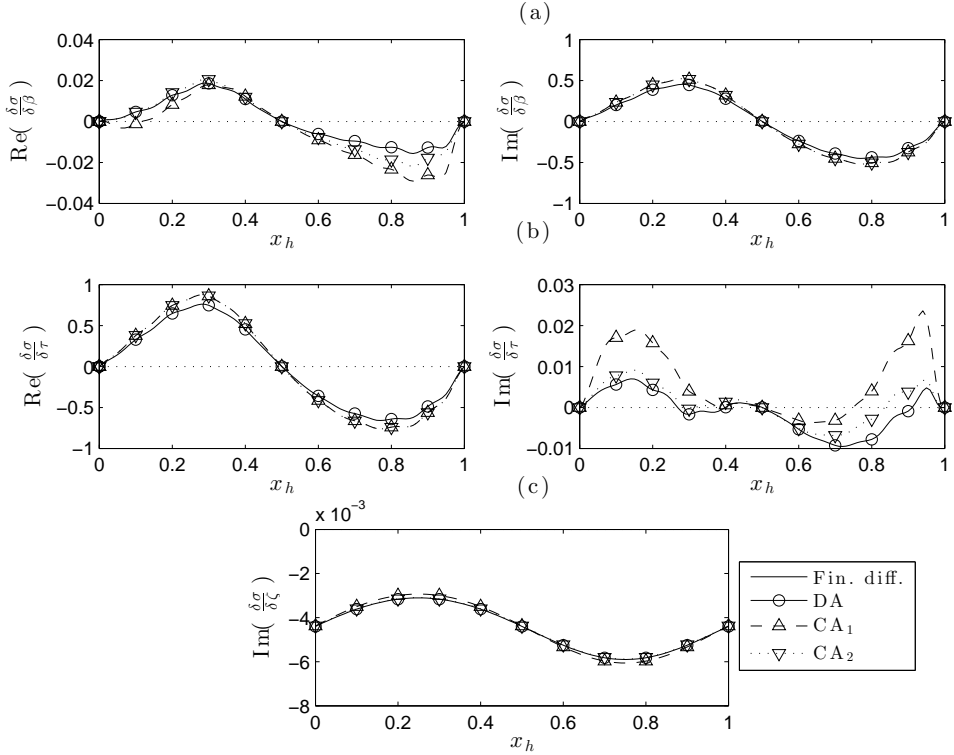}
\caption{Sensitivity to base-state modifications of $\mathit{\beta}$ (a), $\mathit{\tau}$ (b) and $\mathit{\zeta}$ (c). The mean values are $\mathit{\tau}=0.01$,  $\mathit{\beta}=0.5$, $\mathit{c_1}=0.05$ and $\mathit{c_2}=0.005$. For the analysis of $\mathit{\beta}$ and $\mathit{\tau}$ ten Galerkin modes are considered, whereas for $\mathit{\zeta}$ only the first mode is considered.}
\label{fig:3A2}
\end{center}
\end{figure}

Figure \ref{fig:3A2}a shows how a small variation in the heat-release parameter, $\mathit{\beta}$, 
affects the growth rate, $\Real(\mathit{\sigma})$, and the angular frequency, $\Imag\mathit{(\sigma)}$, for different hot-wire positions, $\mathit{x_h}$. 
Figure \ref{fig:3A2}b shows how a small variation in the time-delay coefficient, $\mathit{\tau}$, 
affects the same quantities. 
These are calculated via the $\DA$, $\CAone$, and $\CAtwo$ methods and 
the results are checked against the exact solution, 
which is obtained by finite difference, as in \S \ref{tw}.
\par
As shown in table \ref{tab_base}, these curves depend on the shapes of the direct and adjoint eigenfunctions. In turn, these eigenfunctions are distorted from the natural acoustic modes of the duct by the heat release from the wire. (This distortion is shown in figures \ref{fig:dE} and \ref{fig:aE} for $\mathit{x_h} = 0.25$.) This accounts for the elaborate shapes of the base flow sensitivity curves. 
It is also worth commenting on their relative magnitudes:
small variations in $\mathit{\beta}$ have a much greater effect on the frequency than on the growth rate, 
while small variations in $\mathit{\tau}$ have a much greater effect on the growth rate than on the frequency. 
This will always be the case when $\mathit{\omega \tau} \ll 1$,
which is easy to justify by the following argument.
If $\mathit{p} \sim \sin \omega t$ at the hot wire,
then $\mathit{u} \sim \cos \omega t$ and $\mathit{\dot{q}} \sim \cos \omega(t - \tau)$ there.
Using trigonometric relations,
it is easy to show that 
$\oint \mathit{p} \mathit{\dot{q}} \; \ord \mathit{t}$,
which quantifies how much $\mathit{\beta}$ affects the growth rate, 
is proportional to 
$\sin \mathit{\omega} \mathit{\tau} $
and that
$\oint \mathit{u} \mathit{\dot{q}} \; \ord \mathit{t}$,
which quantifies how much $\mathit{\beta}$ affects the frequency, 
is proportional to
$\cos \mathit{\omega \tau} $.
Therefore, for small $\mathit{\omega\tau}$, 
the change in the growth rate,
$\Real(\mathit{\delta \sigma / \delta \beta})$,
should be of order $\mathit{\omega\tau}$,
while 
the change in the frequency,
$\Imag(\mathit{\delta \sigma / \delta \beta})$,
should be of order $1$.
Differentiating with respect to $\mathit{\tau}$ at constant $\mathit{\beta}$,
we find that 
the change in $\oint \mathit{p} \mathit{\dot{q}} \; \ord \mathit{t}$ due to a change in $\mathit{\tau}$
is proportional to $\mathit{\omega \cos \omega \tau}$. 
Similarly, 
the change in $\mathit{\oint u \dot{q} \; \ord t}$ due to a change in $\mathit{\tau}$ 
is proportional to $\mathit{\omega \sin \omega \tau}$. 
Therefore, for small $\mathit{\omega \tau}$, $\Real(\mathit{\delta \sigma / \delta \tau})$
should be of order $\mathit{\omega}$,
while $\Imag(\mathit{\delta \sigma / \delta \beta})$ 
should be of order $\mathit{\omega}^2 \mathit{\tau}$.
These magnitudes closely match the amplitudes in figure \ref{fig:3A2},
for which $\mathit{\omega} \approx \upi$ and $\mathit{\tau} = 0.01$.

Figure \ref{fig:3A2}c
shows how the angular frequency changes with the damping factor $\mathit{\zeta}$.
A small increase in $\mathit{\zeta}$ lowers the frequency of the linear oscillations. 
A small increase of $\mathit{\zeta}$ is always stabilizing, i.e. the growth rate decreases, but does not depend on the hot-wire position (figure not shown). 
In order to study the sensitivity to small changes of the damping, $\delta\zeta$, only one Galerkin mode has been considered. This is because $\zeta$ is a function of the Galerkin mode, as explained in \S \ref{numdisc}. 
Therefore, with the damping model and numerical discretization adopted, formulae in the bottom row of table \ref{tab_base} are valid only for the first Galerkin mode.

As for the structural sensitivity, 
there is a discrepancy between the $\DA$ and $\CA$ solutions,
which arises from the different truncation errors in the discretizations.
The origin of this error can be inferred from the matrices in appendix A. 
The $\CA_1$ method provides an inaccurate Im($\delta\sigma/\delta\tau$), as shown in figure \ref{fig:3A2}b. 
This is due to the time-delay coefficient and this discrepancy vanishes as the time-delay becomes much smaller.
In this case, we find that the maximal discrepancy between $\CA_1$ and the exact solution is  smaller than 10\% when $\tau<0.001$.
%
%%%%%%%%%%%%%%%%%%%%%%
% FIGURE%
%%%%%%%%%%%%%%%%%%%%%%
%
%
%
\section{Passive control of an unstable system}\label{pas_con} 
\par
%Structural sensitivity analysis via adjoint equations, presented in \S \ref{tw}, can be used to stabilize a thermo-acoustic system. 
In this section we demonstrate the suppression of thermo-acoustic oscillations 
using a control wire placed at the optimal location,
as predicted by the structural sensitivity analysis. 
We use the parameters in figure \ref{fig:2A2}, which shows that, in order to reduce the growth rate most effectively, the control wire should be placed at $\mathit{x_c} = 0.8$.
We integrate the nonlinear time-delayed governing equations (\ref{equ_nl1})-(\ref{equ_nl2}) forward in time with a 4$^{th}$ order Runge-Kutta algorithm and 20 Galerkin modes. 

When the control wire is absent, 
the growth rate is $\mathit{\sigma_r}= 0.00023$ and the angular frequency is $\mathit{\sigma_i}=3.3570$. 
We set the heat-release parameter for the control wire to be $\mathit{\beta_c}=\mathit{\beta}/10=0.05$, which is small enough to fulfil the linear assumptions. 
When the control wire is present,
the growth rate is $\mathit{\sigma_r}=-0.00058$ and the angular frequency is $\mathit{\sigma_i}=3.3354$.
The difference between these values matches
that predicted by the structural sensitivity analysis,
for which $\mathit{\delta \sigma} = \mathit{\beta_c} \times \mathit{\delta\sigma/\delta\beta_c}\approx 0.05 \times (-0.01633-0.4323\mathrm{i}) = -0.00082-0.02162\mathrm{i}$, at $\mathit{x_c}=0.8$.
%
%In this case, the $CA$ methods produce negligible truncation errors
%and are as reliable as the $DA$ method. 

\begin{figure}
\begin{center}
\includegraphics[width=0.99\textwidth, draft = false]{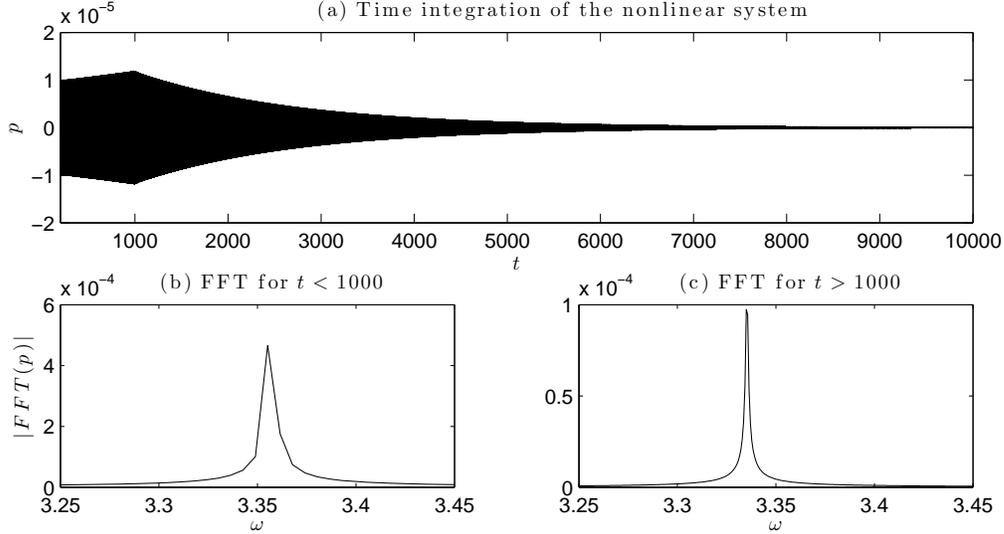}
\caption{Stabilization of the thermo-acoustic system via a second hot wire introduced at $t=1000$ and $\mathit{x_c}=0.8$. $\mathit{\beta_c=\beta}/10=0.05$ and the remaining parameters are the same as in fig. \ref{fig:2A2}. The time integration (a) is performed on the nonlinear time-delayed equations discretized with 20 Galerkin modes. The solution is shown at  $x=0.25$. The amplitude of the spectrum of the solution is shown in (b)-(c).}
\label{fig:time}
\end{center}
\end{figure}

Figure \ref{fig:time}a shows the pressure at $x=0.25$ as a function of time in the nonlinear simulations. The control wire is introduced at $t = 1000$. The behaviour is as expected: there is exponential growth until $\mathit{t} = 1000$ and exponential decay afterwards. In fig. \ref{fig:time}b-\ref{fig:time}c the fast Fourier transform (FFT) performed on the nonlinear time-solution confirms the frequency shift predicted by the sensitivity analysis.
%Note that in fig. \ref{fig:time}b the spectrum is wider than the spectrum of fig. \ref{fig:time}c
%because of the contribution in frequency of the initial transient.
% It is wider because it is taken over a smaller time interval. But this is not important. 

%%%%%%%%%%%%%%%%%%%%%%
% FIGURE%
%%%%%%%%%%%%%%%%%%%%%%
%\input{2009_Rijke_preamble_v2p0}
%\begin{document}
%
\section{Conclusions}
\label{sec_conc}
The main goal of this paper is to take a technique 
developed for the analysis of hydrodynamic stability
and adapt it to the analysis of thermo-acoustic stability.
This technique
uses adjoint equations
to calculate a system's sensitivity to feedback
or to changes in the base state.

By arranging the linearized thermo-acoustic governing equations in two different ways, 
we derive two different sets of adjoint equations,
which we then discretize with a Galerkin decomposition. 
This is known as the `Continuous Adjoint' ($\CA$) method
and the two sets of adjoint equations 
produce two different matrices, labelled $\CAone$ and $\CAtwo$.
We also derive the adjoint equations directly from 
the discretized linearized thermo-acoustic system. 
This is known as the `Discrete Adjoint' ($\DA$) method
and it produces another matrix, labelled $\DA$. 
The $\DA$ matrix is the negative Hermitian 
of the matrix representing the discretized governing equations. 
We calculate the direct and adjoint eigenfunctions of the thermo-acoustic system 
using these direct and adjoint matrices.
We find that the $\DA$ method is more accurate and easier to implement than either $\CA$ method for this thermo-acoustic model.

Two sensitivity analyses are carried out: 
one focuses on structural perturbations 
and the other on base-state perturbations. 
In the structural sensitivity analysis,
we calculate the effect that a generic feedback mechanism has
on the frequency and growth rate of oscillations. 
We illustrate this by considering 
the influence of a second hot wire, with a small heat release parameter.
We find that the second wire affects the frequency much more than the growth rate
and explain this physically by evaluating the Rayleigh Index for the second hot wire.
We then use the results of the structural sensitivity 
to identify the feedback mechanism that is most effective 
at reducing the growth rate of oscillations.
We find that this mechanism should force the momentum equation
in the opposite direction to the velocity perturbation
and that it should be placed at the downstream end of the duct.
An adiabatic fine mesh would achieve this. 
%%
%We find that this control wire has a stabilising effect 
%only when placed well into the second half of the tube,
%and not in the entire second half, as might be expected a priori.
%We explain this by ... 
%
In the base-state sensitivity analysis,
we calculate the effect that a small variation in the base-flow parameters has
on the frequency and growth rate of oscillations. 
As expected,
we find that a small increase in the wire temperature affects the frequency more than the growth rate
and that a small increase in the time delay affects the growth rate more than the frequency. 
Also as expected, 
we find that a small increase in the damping
always has a stabilizing effect. 
The novelty of this paper is % not so much in the results,
%which are well known, but 
in the technique.
Each sensitivity analysis was obtained extremely quickly with a single calculation.
It was then checked against the exact solution 
found by many finite difference calculations.
The $\DA$ method matched the finite difference method exactly,
while there was some discrepancy when using the $\CAone$ and $\CAtwo$ methods.

The successful application of 
sensitivity analysis to thermo-acoustics 
opens up new possibilities for the passive control of thermo-acoustic oscillations.
In a single calculation, 
sensitivity analysis shows
how the growth rate and frequency of small oscillations about some base state
are affected
either by a passive control element in the system
or by a change to its base state. 
This gradient information
can be combined with other constraints,
such as that the total mean heat release be constant,
to show how an unstable thermo-acoustic system 
should be changed in order to make it stable. 
In this paper, 
we have demonstrated this for a simple system 
with four elements to the base state:
the hot wire position, its heat-release coefficient, its time delay and the damping.
In future work,
we will 
examine more elaborate flame models and acoustic networks. 
This will allow us to 
calculate the sensitivity to the flame shape
and to the characteristics of the acoustic network in which the flame sits.
\newline
\par
We would like to thank Dr. Outi Tammisola (University of Cambridge, Department of Engineering, U.K.) for valuable discussions and comments on this paper. 
This work was supported by the European Research Council through Project ALORS 2590620.

\appendix
\section{Discretized equations} \label{append}
It is useful to define 
the following matrices,
which are expressed in matrix notation (repeated indices are not to be summed):
\begin{equation}
\mathsfbi{A_{ij}}\equiv0,\;\;\;\mathsfbi{B_{ij}}\equiv\upi\delta_{ij}i,\;\;\;\mathsfbi{E_{ij}}(c_{1},c_{2})\equiv-\zeta_i\delta_{ij}, \nonumber
\end{equation}
\begin{equation}
\mathsfbi{F_{ij}}(\beta_w,x_w)\equiv -2\beta_w \sin(\upi i x_w)\cos(\upi j x_w), \;\;\;\mathsfbi{G_{ij}}(\beta_w,x_w,\tau_w)\equiv 2i\upi\tau_w\beta_w  \sin(\upi i x_w)\cos(\upi j x_w),\nonumber\end{equation}
\begin{equation}
\mathsfbi{H_{ij}}(\beta_w,x_w,\tau_w,c_{1},c_{2}) \equiv 2\beta_w\tau_w\zeta_j \cos(\upi i x_w)\sin(\upi j x_w),\nonumber
\end{equation}
\begin{equation}
\mathsfbi{C_{ij}}(\beta_w,x_w)\equiv -\mathsfbi{B_{ij}}+\mathsfbi{F_{ij}},\;\;\;\mathsfbi{D_{ij}}(\beta_w,x_w,\tau_w,c_{1},c_{2})\equiv \mathsfbi{E_{ij}}+\mathsfbi{G_{ij}}. \nonumber
\end{equation}
where $i, j=1,2,...,N$, $N$ is the number of Galerkin modes, $\mathit{\delta_{ij}}$ is the Kronecker delta and $\mathit{w}$ stands for \emph{wire}. The direct matrix $\mathsfbi{\Gamma}$ is given by:
\begin{equation}
\mathsfbi{\Gamma} =\left[\begin{array} {cc} \mathsfbi{A} & \mathsfbi{B}\\ \mathsfbi{C}(\beta_h, x_h) & \;\;\;\;\; \mathsfbi{D}(\beta_h,x_h,\tau_h,c_{1},c_{2})  \end{array} \right] \label{gammaAPP}.
\end{equation}

The continuous adjoint equations (table \ref{tab_adj}) are discretized using the Galerkin method as for the direct modes, by means of the decomposition in eq. (\ref{equ_gal_u}). The discretization of the first set of adjoint equation $\CAone$ (table \ref{tab_adj}) gives rise  to the following adjoint matrix
\begin{equation}
\mathsfbi{\Phi} =\left[\begin{array} {cc} -\mathsfbi{G}^T(\beta_h, x_h, \tau_h) &\;\;\;\;\mathsfbi{B}-\mathsfbi{F}^T(\beta_h, x_h)+\mathsfbi{H}(\beta_h,x_h,\tau_h,c_{1},c_{2})\\ -\mathsfbi{B} & -\mathsfbi{E}(c_{1},c_{2}) \end{array} \right] , \label{phi1}
\end{equation}
while the second set $\CAtwo$ (table \ref{tab_adj}) gives the following adjoint matrix 
\begin{equation}
\mathsfbi{\Phi} =\left[\begin{array} {cc} \mathsfbi{A} & \mathsfbi{B}-\mathsfbi{F}^T(\beta_h, x_h)\\ -\mathsfbi{B} & \;\;\;\;-\mathsfbi{E}(c_{1},c_{2})+\mathsfbi{G}^T(\beta_h, x_h,\tau_h)  \end{array} \right] \label{phi2}.
\end{equation}  
Note that $\mathsfbi{\Gamma}$ and $\mathsfbi{\Phi}$ are $2\mathit{N}\times2\mathit{N}$ matrices. We indicated the main hot wire with subscript $h$ and the control hot wire with the subscript $c$.
Finally, the perturbation matrix of the direct system is:
\begin{equation}
\mathsfbi{\delta P} =\left[ \begin{array} {cccc} \Large{[0]}_{N\times N} & \;\;\;\;\Large{[0]}_{N\times N}\\ \\ \mathsfbi{C}(\beta_h+\delta\beta_h, x_h) +\ldots & \;\;\;\;\mathsfbi{D}(\beta_h+\delta\beta_h, x_h, \tau_h+\delta\tau_h,c_1+\delta c_{1},c_2+\delta c_{2})+\ldots \\ \ldots +\mathsfbi{C}(\delta\beta_c, x_c)& \;\;\;\;\ldots+\mathsfbi{D}(\delta\beta_c, x_c, \delta\tau_c,c_1,c_2)  \end{array} \right] . \label{we}
\end{equation}
On the one hand,
we obtain the perturbation matrix caused by the presence of the second hot wire 
by setting $\mathit{\delta\beta_h}=\mathit{\delta\tau_h}=\mathit{\delta c_{1}}=\mathit{\delta c_{2}}=0$ and $\mathit{\delta\beta_c}>0$ and $\mathit{\delta\tau_c}>0$.
On the other hand,
we obtain the perturbation matrix caused by (positive) base-flow variations
by setting  $\mathit{\delta\beta_c}=\mathit{\delta\tau_c}=0$ and $\mathit{\delta\beta_h}>0$, $\mathit{\delta\tau_h}>0$, $\mathit{\delta c_{1}}>0$ and $\mathit{\delta c_{2}}>0$.

\section{Nonlinear time-delayed equations for control} \label{append2}
In this section we provide the nonlinear time-delayed equations of the thermo-acoustic system with a control hot wire. 

Referring to the time integration presented in \S \ref{pas_con}, when the second hot wire is off, for $t<1000$, then $\mathit{\beta_c}=0$; when the second wire is on, for $t\geq1000$, then $\mathit{\beta_c}=\mathit{\beta}/10$.  
\begin{equation}
\label{equ_nl1}
\frac{\partial u}{\partial t} + \frac{\partial p}{\partial x} = 0 ,
\end{equation}
\begin{equation}
\label{equ_nl2}
\frac{\partial p}{\partial t} + \frac{\partial u}{\partial x}
+ \zeta p
- \frac{2}{\sqrt{3}}\beta
\left(
\left|
\frac{1}{3} + u(t - \tau)
\right|^{\frac{1}{2}}
-
\left(
\frac{1}{3}
\right)^{\frac{1}{2}}
\right)
\delta(x - x_h) + \ldots
\end{equation}
\begin{equation}
\ldots- \frac{2}{\sqrt{3}}\beta_c
\left(
\left|
\frac{1}{3} + u(t - \tau_c)
\right|^{\frac{1}{2}}
-
\left(
\frac{1}{3}
\right)^{\frac{1}{2}}
\right)
\delta(x - x_c) =0. \nonumber
\end{equation}

% References
%

\end{document}